\documentclass[pre,twocolumn]{revtex4-1}
\usepackage{amsmath,amssymb,latexsym,epsfig,graphics,epsf}
\usepackage{graphics,xcolor}
\usepackage{float}
\newcommand{\tD}{\tilde{D}}
\newcommand{\tom}{\tilde{\omega}}
\newcommand{\bR}{\bold R}

\newcommand{\be}{\begin{equation}}
\newcommand{\ee}{\end{equation}}
\newcommand{\fig}[1]{Fig.~\ref{#1}}
\newcommand{\Fig}[1]{Figure~\ref{#1}}
\newcommand{\eq}[1]{Eq.~(\ref{#1})}
\newcommand{\Eq}[1]{Equation~(\ref{#1})}

\begin{document}
\title{Solid-like mean-square displacement in  glass-forming liquids}
\author{Thomas B. Schr{\o}der and Jeppe C. Dyre}
\affiliation{"Glass and Time", IMFUFA, Dept. of Science and Environment, Roskilde University, P. O. Box 260, DK-4000 Roskilde, Denmark}

\date{\today}

\begin{abstract}  
It was recently shown that the real part of the frequency-dependent fluidity for several glass-forming liquids of different chemistry conforms to the prediction of the random barrier model (RBM) devised for ac electrical conduction in disordered solids [S. P. Bierwirth \textit{et al.}, Phys. Rev. Lett. {\bf 119}, 248001 (2017)]. Inspired by these results we introduce a crystallization-resistant modification of the Kob-Andersen binary Lennard-Jones mixture for which the results of extensive graphics-processing unit (GPU)-based molecular-dynamics simulations are presented. We find that the low-temperature mean-square displacement is fitted well by the RBM prediction, which involves no shape parameters. This finding highlights the challenge of explaining why a simple model based on hopping of non-interacting particles in a fixed random energy landscape can reproduce the complex and highly cooperative dynamics of glass-forming liquids. 
\end{abstract}

\maketitle

From experimental data for nine glass-forming liquids, Gainaru and coworkers recently demonstrated a striking universality of the real part of the frequency-dependent fluidity \cite{bie17}. The fluidity is defined as $1/\eta(\omega)$ in which $\omega$ is the angular frequency and $\eta(\omega)$ the complex frequency-dependent linear shear viscosity. The data involved van der Waals, ionic, and hydrogen-bonding liquids, i.e., chemically quite diverse systems. The universal fluidity data were shown to fit well to a prediction based on the random barrier model (RBM). This is surprising in view of the fact that this model has no shape parameters and was devised for describing the completely different hopping conduction in disordered solids. The RBM considers non-interacting particles jumping stochastically on a simple cubic lattice with identical site energies and random energy barriers for nearest-neighbor jumps \cite{dyr85,dyr88,sch00}, which is not at all how one thinks about a liquid. To illuminate this puzzling situation we have carried out extensive computer simulations of a highly viscous model liquid in order to see whether the RBM does describe the particle dynamics.

The relaxation time increases dramatically when a liquid is supercooled and approaches the glass transition \cite{bra85,ang95,deb01,dyr06,ber11}. A standard probe of the dynamics is the single-particle mean-square displacement (MSD) as a function of time, $\langle\Delta r^2(t)\rangle$, in which $\Delta r(t)$ is the distance traveled by a given atom or molecule in time $t$ and the angular brackets denote an ensemble average. At long times, the MSD is proportional to time and determines the (self) diffusion coefficient $D$ via $\langle\Delta r^2(t)\rangle = 6Dt$. The transition to a linear-time MSD takes place roughly at the time at which the particles on average have moved an interatomic distance. Since all liquids become diffusive at long times, it is the  subdiffusive regime that reveals details about the liquid dynamics. 

The RBM was devised as an idealized model of ac ionic or electronic hopping conduction in disordered solids like oxide glasses, polymers, amorphous semiconductors, etc \cite{dyr85,dyr88}. In the extreme-disorder limit, i.e., when $k_BT$ is much smaller than the relevant energy barriers, the model predicts a universal MSD such that the MSD as a function of time is the same for all barrier distributions except for a scaling of time and space \cite{sch00}. The only requirement for this to apply is that the energy barrier probability distribution is continuous at the percolation threshold \cite{sch00}. Physically, the response is universal because the dynamics at extreme disorder is dominated by percolation in the 3d random energy landscape \cite{sch00}. 

In a simple analytical approximation the frequency-dependent diffusion coefficient of the RBM defined by $D(\omega)\equiv (-\omega^2/6)\int_{0}^{\infty}\langle\Delta r^2(t)\rangle\exp(i\omega t)dt$ is predicted to be the solution of $\ln\tD=(i\tom/\tD)^{2/3}$ in which $\tom$ is a scaled frequency and $\tD\equiv D(\omega)/D(0)$ \cite{sch08}. This is derived by combining the Alexander-Orbach conjecture that the percolation cluster has harmonic dimension $4/3$ (independent of dimension) \cite{ale82} with the effective-medium approximation applied to diffusion on the percolation cluster \cite{sch08}. The quoted equation provides an excellent fit to computer simulations of the RBM \cite{sch08}, except at the lowest frequencies where the transition to a  frequency-independent diffusion constant is better described by the solution of the following equation $\ln\tD=(i\tom/\tD)(1+(8/3)i\tom/\tD)^{-1/3}$ \cite{sch08}.

MSD simulation data are conveniently fitted to the von Schweidler empirical expression \cite{ka1}
\be\label{vSch}
\langle\Delta r^2(t)\rangle
\,=\, r_0^2\,+\,a(6Dt)^b\,+\,6Dt\,.
\ee
According to mode-coupling theory, the exponent $b$ is non-universal \cite{got08}. Tokuyama has discussed common features of the MSD of different models \cite{tok11}, but to the best or our knowledge the possibility of a universal viscous-liquid MSD has not been considered in the literature. This means that after the publication of Ref. \onlinecite{bie17}, the glass community finds itself in the unusual situation that experiments suggest a more universal behavior than reported in simulations. An important difference between experiments and simulations, however, is that the latter cannot yet cover the long time scales of experiments on highly viscous liquids. Is this why the MSDs reported in simulations, though similar, are not universal? To address this question one needs a viscous model liquid that is fast and easy to simulate and which does not crystallize, even in extremely long simulations. 

Recent exciting developments with swap dynamics have made it possible to generate equilibrium states of liquids with astronomically long relaxation times \cite{nin17,oza19}. Unfortunately, probing the alpha relaxation dynamics on these time scales remains out of reach, so for studying the equilibrium dynamics brute-force molecular dynamics (MD) is still the only available option. We utilize state-of-art graphics-processing unit (GPU) simulations \cite{RUMD} to access equilibrium dynamics at very low temperatures. The duration of the longest simulation was four months, which with traditional CPU computing would have taken several years. 

For almost a century the standard model in liquid-state theory has been the Lennard-Jones (LJ) system based on the following pair potential in which $\varepsilon$ is a characteristic energy and $\sigma$ a characteristic length: $v(r)=4\varepsilon\left[\left({r}/{\sigma}\right)^{-12}-\left({r}/{\sigma}\right)^{-6}\right]$ \cite{lj24,han13}. The LJ liquid cannot be studied in the supercooled phase because it crystallizes. In 1995 Kob and Andersen proposed a binary LJ system that is easily supercooled. The Kob-Andersen (KA) model is a mixture of 80\% large A particles and 20\% small B particles \cite{ka1}. The trick is to have a strong AB non-ideal (non-Lorentz-Berthelot) attraction impeding phase separation. The parameters of the KA model are \cite{ka1} $\sigma_{\rm BB}/\sigma_{\rm AA}=0.88$, $\sigma_{\rm AB}/\sigma_{\rm AA}=0.8$, $\varepsilon_{\rm AB}/\varepsilon_{\rm AA}=1.5$, and $\varepsilon_{\rm BB}/\varepsilon_{\rm AA}=0.5$. The KA model quickly became the standard model for simulations of viscous liquid dynamics \cite{ash03}. The mode-coupling temperature (the temperature at which idealized mode-coupling theory based on higher-temperature data predicts a diverging relaxation time \cite{oza19}) was estimated to be $T_c=0.435$ \cite{ka1}. As computers became faster, it eventually became possible to investigate the model below $T_c$, see e.g. Refs. \onlinecite{ash03} and \onlinecite{cos18a}. The KA model crystallizes in very lengthy simulations \cite{tox09,ing18a}; in fact, at the standard density 1.2 the KA liquid is supercooled whenever $T<T_m=1.03$ \cite{ped18}. Although the strong AB attraction impedes phase separation, the supercooled KA system eventually does crystallize by phase separating into a pure A phase \cite{ped18}.

\begin{figure}[htbp!]
	\includegraphics[width=8cm]{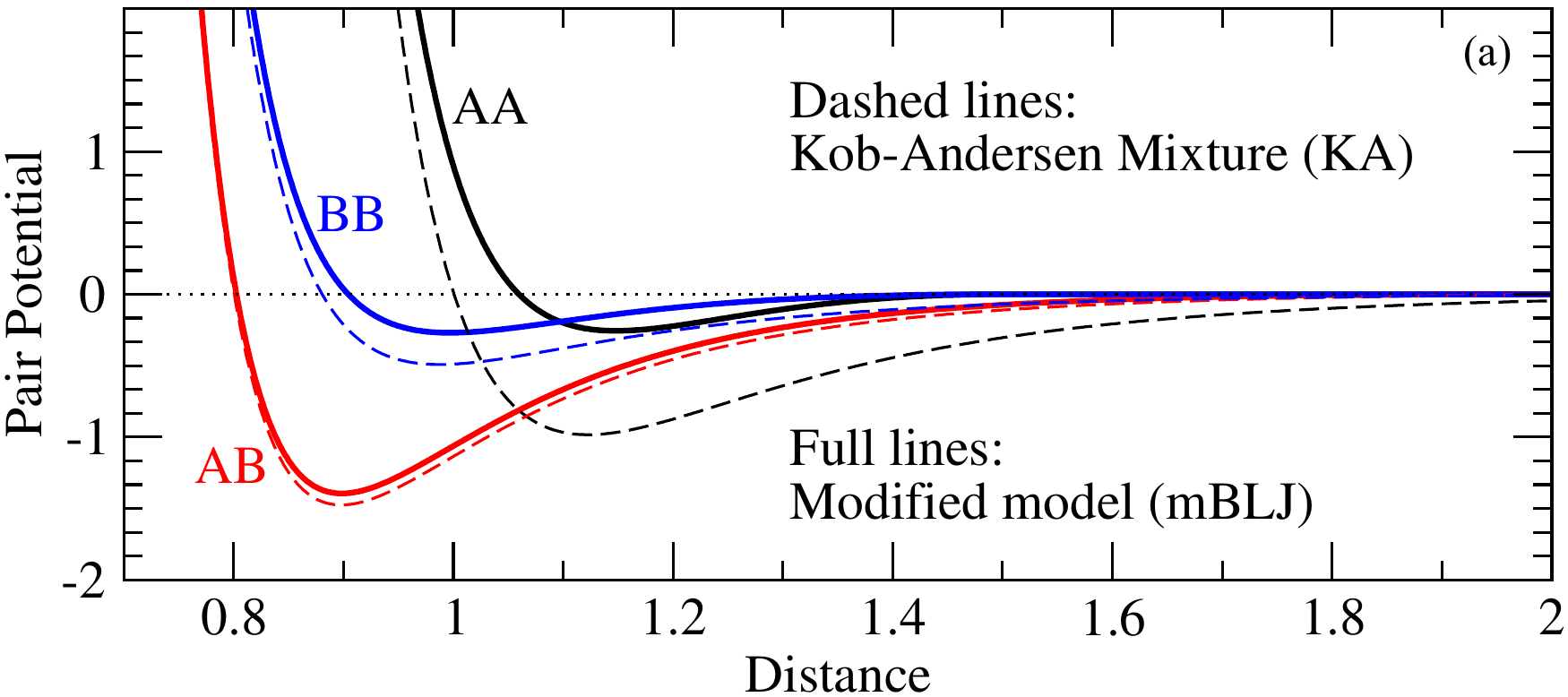}
	\includegraphics[width=8cm]{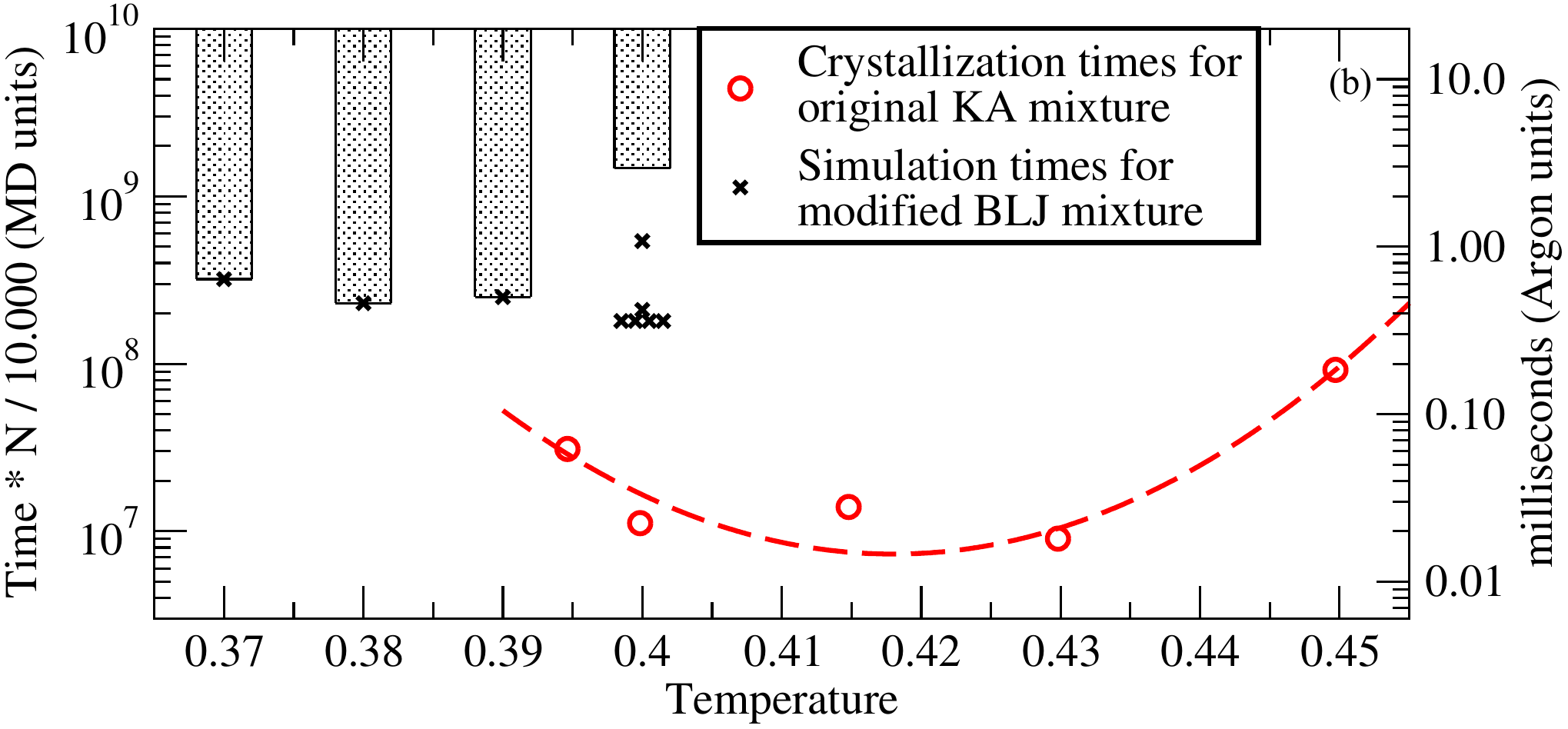}
	\caption{\label{fig1}
        (a) Pair potentials in units of $\varepsilon_{\rm AA}$ plotted as a function of pair distance in units of $\sigma_{\rm AA}$. Dashed lines: The standard Kob-Andersen (KA) pair potential with shifted potential cutoffs at  $r=2.5\,\sigma_{\rm\alpha\beta}$. Full lines: The modified binary Lennard-Jones (mBLJ) pair potentials, which introduce shifted-force cutoffs at $r=1.5\,\sigma_{\rm AA}$ for the AA and BB interactions and at $r=2.5\,\sigma_{\rm AB}$ for the AB interaction. The significant reduction of the AA attraction obtained in this way suppresses the tendency for phase separation.
        (b) The red circles give the average crystallization times of the KA model \cite{ing18a}, the red dashed curve is a parabolic fit to these data. The figure also shows the simulation times for the mBLJ model at four temperatures: $T=0.37;\,0.38;\,0.39;\,0.40$ (black crosses). Simulation times are scaled to be comparable to the crystallization times of Ref. \onlinecite{ing18a}, which used $N=10000$.
        At $T=0.40$ several independent simulations were performed, none of which crystallized. From this fact one deduces a higher estimate of the minimum crystallization time. At each temperature the black rectangles indicate the estimated range of possible crystallization times for the mBLJ model.}
\end{figure} 

Is it possible to modify the KA model to make it even less prone to crystallization? We do this by introducing a shifted-force cutoff at $r=1.5\,\sigma_{\rm AA}$ for the AA and BB interactions \cite{tildesley,tox11a}. \Fig{fig1}(a) shows the original KA pair potentials (dashed lines) compared to the modified binary Lennard-Jones (mBLJ) pair potentials (full lines). The AA attraction in the later is visibly weaker and the BB attraction has also been weakened. The motivation for using a shifted-force cutoff is that this is known to leave the liquid dynamics almost unchanged \cite{tildesley,tox11a}, thus facilitating a comparison between the original and the modified model.

We performed molecular-dynamics simulations in the $NVT$ ensemble with $N=8000$ particles (unless otherwise noted) at the ``standard'' number density $\rho\equiv N/V = 1.2$; the temperature $T$ was controlled by a Nose-Hoover thermostat.  Unless otherwise noted, results are reported in standard MD-units defined by $\sigma_{AA} = 1$,  $\epsilon_{AA}=1$, $m_A=m_B=1$, and $k_B=1$. The time step was $0.005$ in this unit system.

The mBLJ liquid did not crystallize during the months of GPU simulations performed for this study. Figure \ref{fig1}(b) shows as a function of temperature the average crystallization time of the original KA model (red circles) \cite{ing18a}, and the total simulation times for the mBLJ system (black crosses). Since the mBLJ system did not crystallize, at each temperature the total simulation time gives a lower bound on the crystallization time. The black rectangles indicate where the unknown crystallization times are to be found. At $T=0.40$ several independent simulations were performed. This includes simulations that were first equilibrated at the lower temperatures (0.37, 0.38, and 0.39, respectively), a procedure known to increase the tendency to  crystallize. Based on the data presented, we estimate that the mBLJ liquid has at least a 100 times longer crystallization time than the original KA liquid.

\begin{figure}[htbp!]
	\includegraphics[width=8cm]{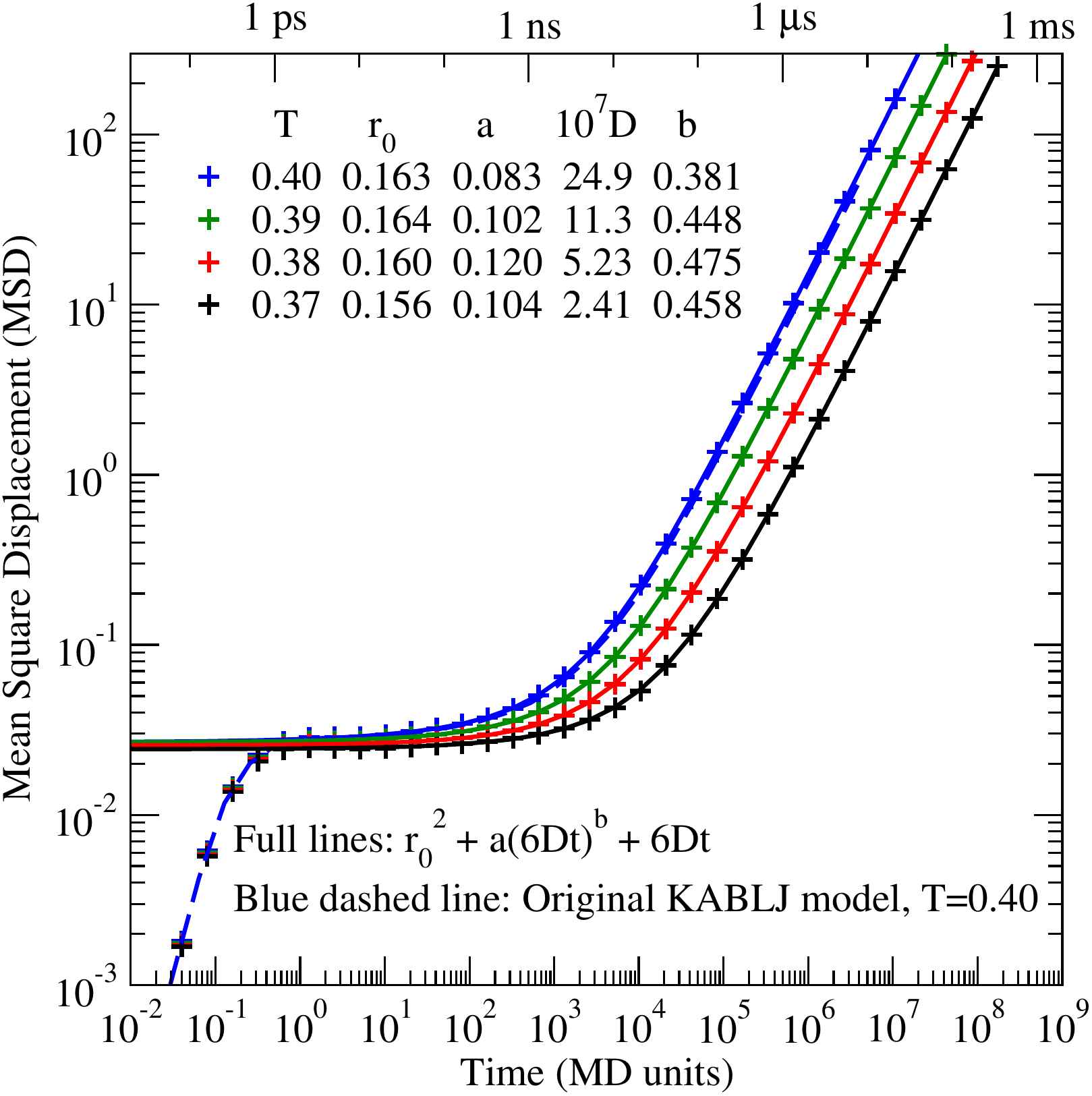}
	\caption{\label{fig2}
All-particle MSD as a function of time for the mBLJ system at the four temperatures $T=0.37;\,0.38;\,0.39;\,0.40$. Full lines are best fit to the von Schweidler expression \eq{vSch} \cite{ka1}.  At short times the data follow the line of slope two expected from ballistic behavior, at long times the slope is unity reflecting diffusive behavior. The liquid dynamics of the mBLJ model is close to that of the KA model, which is illustrated by the dashed blue line giving the MSD of the KA model at $T=0.40$.
}
\end{figure} 

Having modified the original KA model such that crystallization is in practice avoided, we turn to studying the supercooled liquid dynamics. \Fig{fig2} shows the mBLJ liquid's all-particle MSD as a function of time at four temperatures. The figure presents data going to times larger than $10^8$ MD time units, corresponding to 0.2 milliseconds in argon units. The data are for $N=8000$ particles; size independence was checked by simulating also $N=27000$ particles at $T=0.40$, which gave indistinguishable results. The dashed blue line gives data for the original KA model at $T=0.40$, which are close to those of the mBLJ model (blue crosses). This confirms that the two models have very similar liquid dynamics.

At very short times one finds the well-known free-particle ballistic MSD $\propto t^2$ after which there is a plateau where the MSD is almost constant. This derives from ``cage rattling''  of the particles in local potential-energy minima, reflecting the fact that a viscous liquid over short time scales is virtually indistinguishable from an amorphous solid. At longer times the MSD increases, of course, and eventually one finds the standard diffusive behavior MSD $\propto t$. Note the dramatic slowing down upon cooling: a temperature decrease of 7.5\% results in more than one decade's slowing down. What causes this is, in a nutshell, the mystery of the glass transition \cite{dyr06,ber11}. The full curves in \fig{fig2} are fits to the von Schweidler expression \eq{vSch}.

\begin{figure}[htbp!]
	\includegraphics[width=8cm]{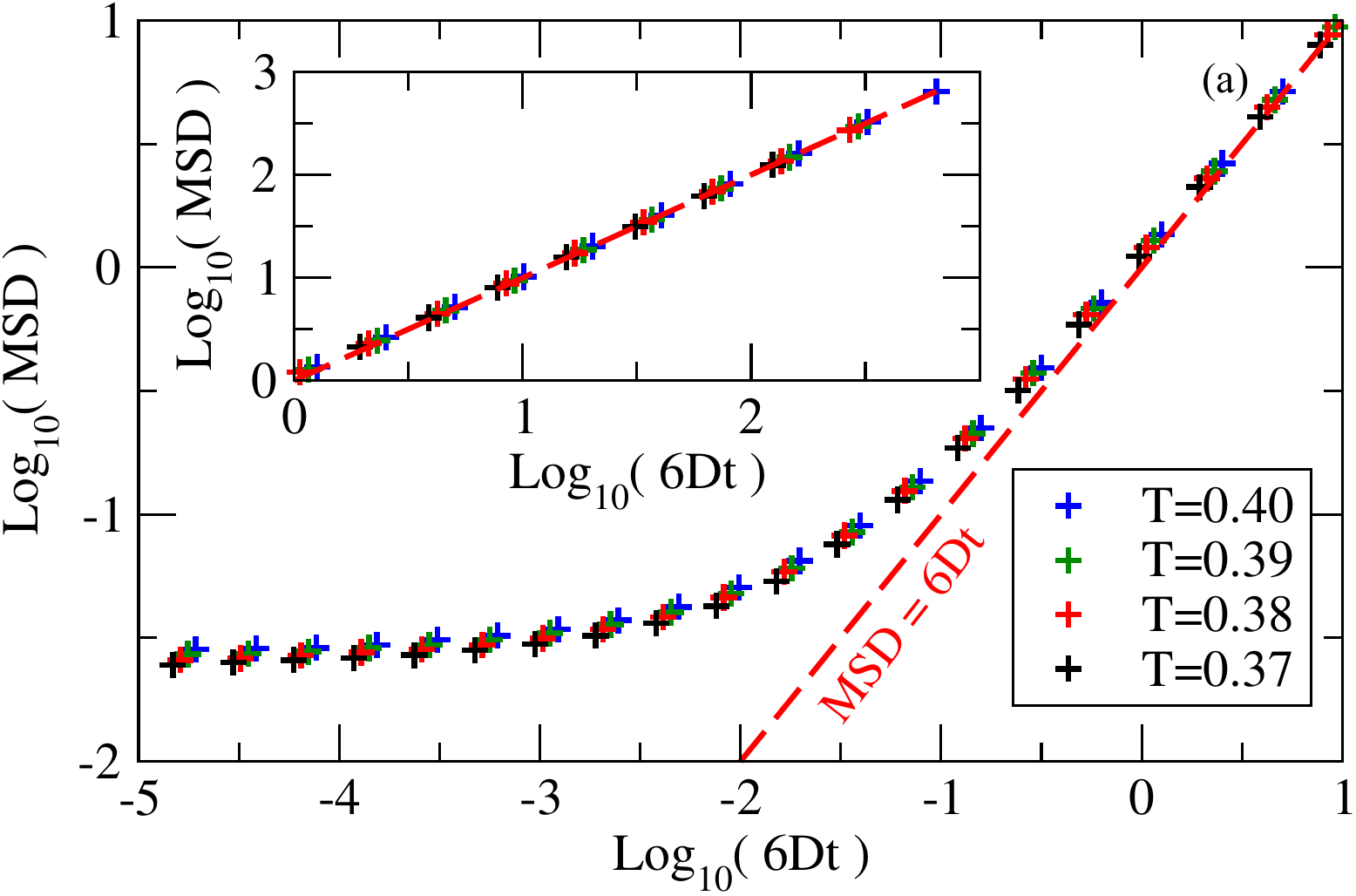}
	\includegraphics[width=8cm]{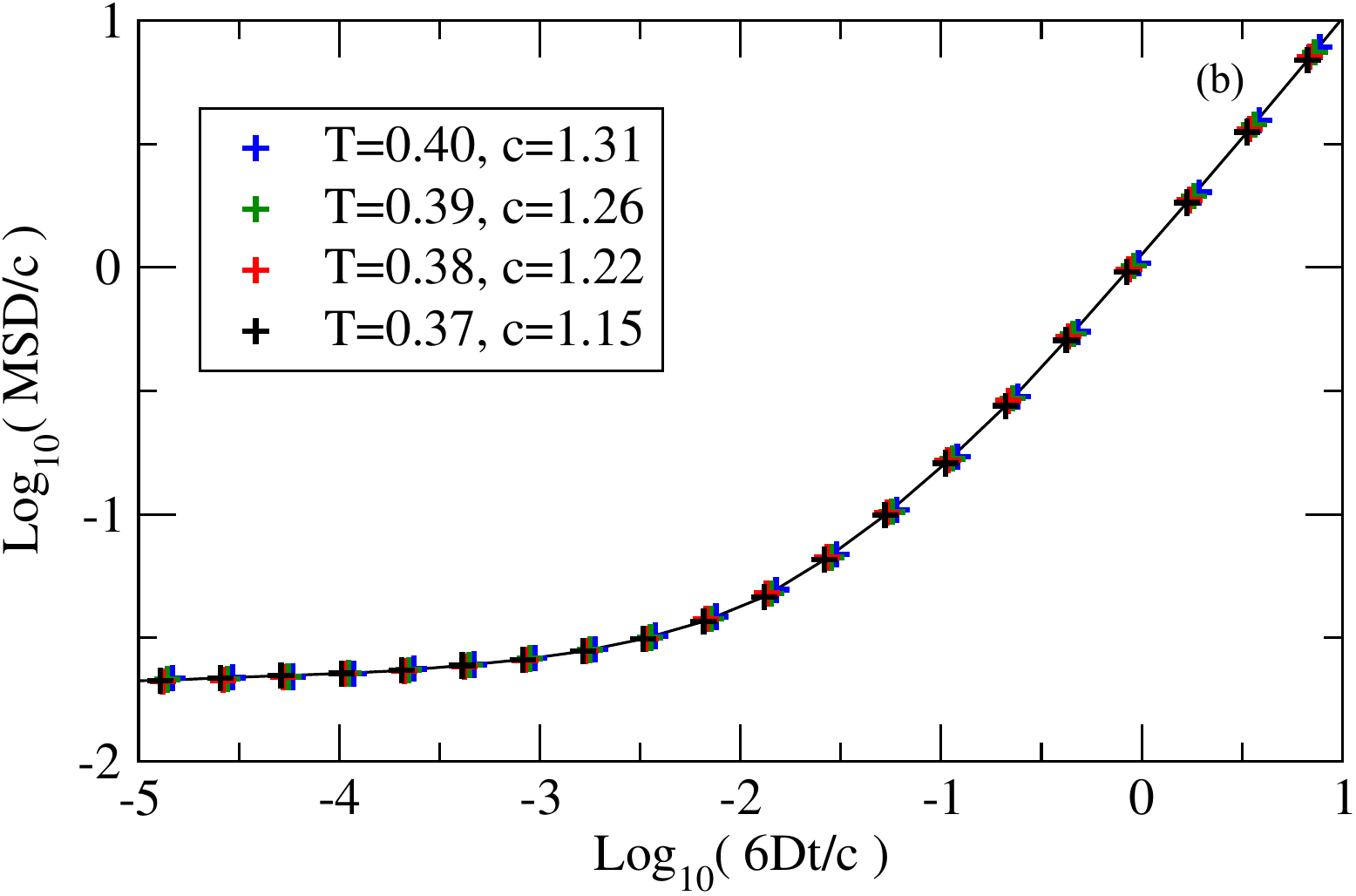}
	\caption{\label{fig2.2}	
    (a) All-particle MSD at the four temperatures plotted as a function of time scaled by the diffusion coefficient $D$, showing data corresponding to $t>20$ in MD units. The data superpose not just at long times, but also in the transition region. The short-time ``plateau'' regions, however, change with temperature. The inset shows results for longer times, demonstrating that the diffusive behavior is given by the red dashed curve.
	(b) Same data as in (a) but scaled further on both axes by a squared empirical length $c$, showing a near-perfect collapse, i.e., demonstrating time-temperature superposition. The black line is the von Schweidler fit to the the $T=0.40$ data. }
\end{figure} 

\begin{figure}[htbp!]
	\includegraphics[width=8cm]{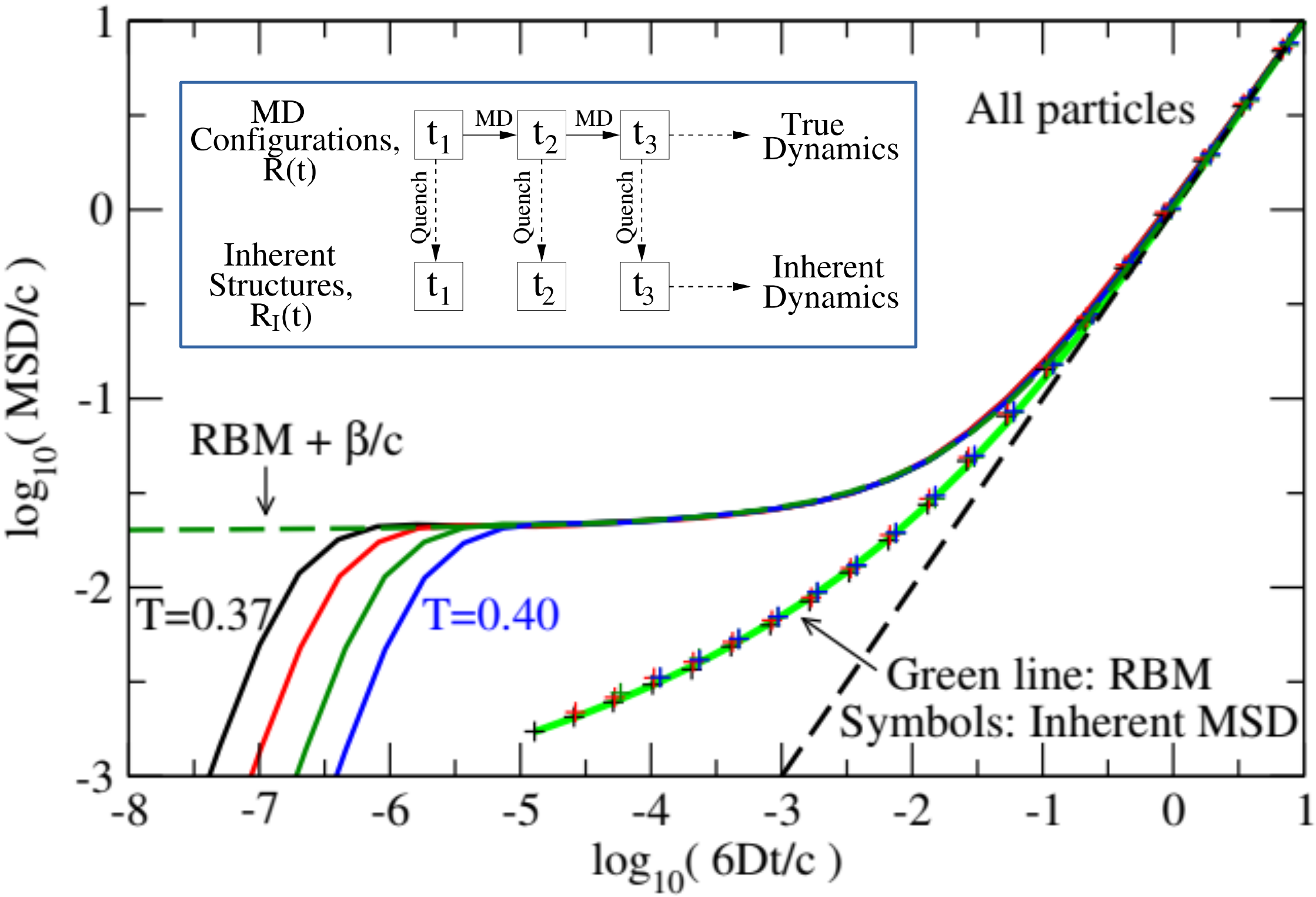}
	\caption{\label{fig3}
	All-particle MSD of true and inherent dynamics (full lines and plus symbols) for four temperatures compared to the RBM prediction, \eq{fiteq} and \eq{fiteq_inherent} respectively. The fit parameters were determined as described in the text. The agreement with the inherent dynamics demonstrates that the RBM, despite having no shape parameters, provides a good representation of the liquid dynamics. Adding a constant to represent the contribution from cage-rattling gives very good agreement with the true MSD.
}
\end{figure}

\Fig{fig2.2} investigates time-temperature superposition (TTS), i.e., whether data for different temperatures can be made to collapse by scaling of the axes, as found in many experiments \cite{bie17}. In their original paper, Kob and Andersen reported TTS obtained by scaling time with the diffusion coefficient. In \fig{fig2.2}(a) we apply the same scaling to our low-temperature MSD data, finding differences in the plateau regime. In \fig{fig2.2}(b) we perform a further scaling on both axes by a parameter $c$ that has dimension of a squared length, thus making both axes dimensionless. The determination of $c$ is described below in connection with  \fig{fig3}. From \fig{fig2.2}(b) one concludes that TTS applies. 

Next we compare our low-temperature MSD data to the RBM prediction (\fig{fig3}). Full lines are the MSD data scaled as in \fig{fig2.2}, including now also the ballistic regime. The RBM relates to spatially discrete particle jumps, so the predicted MSD does not include the short-time ``cage-rattling'' plateau present in any realistic viscous liquid model. Consequently, in order to compare the RBM to the MSD data we add a constant reflecting the cage-rattling contribution to the MSD. If the universal RBM prediction for the MSD corresponding to unit diffusion coefficient is denoted by $F_{\rm RBM}(t)$, the RBM prediction thus becomes: 

\be 
  \langle\Delta r^2(t)\rangle=c\, F_{\rm RBM}(\alpha t)\,+\,\beta  \label{fiteq} 
\ee
with three parameters $c,\,\alpha,\,\beta$. For dimensional reasons, there must be at least two parameters, a length and a time. The long-time limit $\langle\Delta r^2(t)\rangle=6Dt$ results in $c\alpha=D$. \Eq{fiteq} is plotted as the green dashed line in \fig{fig3}. We conclude that, despite having just a single shape parameter -- compared to the two shape parameters of the von Schweidler expression (\eq{vSch}) -- \eq{fiteq} fits the MSD data very well.

In the following we show that one can go one step further and compare the RBM model to the dynamics of the mBLJ model using just the two scaling parameters $D$ and $c$, i.e., without any shape parameters. Already in 1969 Goldstein recognized the significance of potential-energy minima \cite{gol69}, which were later termed ``inherent states'' by Stillinger and Weber \cite{sti83}. If the $N$ particle coordinates are collected into a single $3N$-dimensional vector denoted by $\bR$, one can distinguish between  the ``true'' Newtonian dynamics $\bR(t)$ and its corresponding quenched ``inherent'' dynamics $\bR_{\rm I}(t)$. As illustrated in the inset of \fig{fig3}, the latter is arrived at by quenching configurations from an equilibrium MD simulation to their inherent states \cite{sch00a}. We run the same data analysis program on both the true configurations $\bR(t)$ and the  quenched  inherent configurations $\bR_{\rm I}(t)$. Note that $\bR_{\rm I}(t)$ in the course of time jumps discontinuously from one constant vector to another. Below $T_c$ the dynamics separates into  oscillations around inherent states and transitions between these \cite{sch00a}, as predicted by Goldstein \cite{gol69}. The point is that the effect of oscillations is removed by considering the inherent dynamics.  Thus the inherent MSD, $\langle\Delta r_I^2(t)\rangle$, can be compared directly to the RBM prediction without additional constants:

\be
  \langle\Delta r_I^2(t)\rangle=c\, F_{\rm RBM}(\alpha t)\,  \label{fiteq_inherent}\,.
\ee
In \fig{fig3} the inherent MSD is plotted as crosses for all four temperatures. It obeys time-temperature superposition and, despite having no shape parameters, the RBM prediction \eq{fiteq_inherent} (full green line) agrees nicely with the data.

In the above analysis, the scaling parameter $c$ was determined by minimizing the root-mean-square difference between the inherent MSD and the RBM prediction (\eq{fiteq_inherent}). Subsequently, the plateau  parameter $\beta$ was fitted for the true MSD (\eq{fiteq}). The true MSD can be fitted directly to \eq{fiteq}, but using the resulting $c$ parameters for the inherent MSD results in a considerably worse fit. This reflects the fact that the inherent contribution to the MSD is small at short times. Note, however, that the inherent MSD at short times is more than a factor 100 \emph{larger} than the extrapolation of the diffusive regime (black dashed line). This is comparable to the increase in fluidity observed in experimental data \cite{bie17}. 

We turn now to the relation between our results and the experimental findings. If the RBM describes the liquid MSD and if one assumes that the macroscopic shear viscosity controls the microscopic frictional forces via the Stokes-Einstein relation between diffusion coefficient and viscosity, the frequency-dependent fluidity is proportional to the RBM universal prediction as found for nine liquids by Gainaru and coworkers \cite{bie17}. Both assumptions are highly nontrivial. The first assumption is supported by our simulation results presented above. The second assumption, the Stokes-Einstein assumption, is definitely challenged in glass-forming liquids \cite{hod93,tar95,dou98,deb01,sen14}. We shall not discuss this further here, but note that for the arguments presented it is enough that the frequency-dependent viscosity and diffusivity are inversely proportional -- the proportionality coefficient may well depend on temperature.

Why was the plateau parameter $\beta$ not necessary in the analysis of experimental data in Ref. \onlinecite{bie17}? Letting tilde denote a suitable scaling, we get from \eq{fiteq}:

\be 
  \tilde{F}( \tilde\omega) = \tilde F_{\rm RBM}(\tilde\omega)\,+ i\tilde\omega\beta  \label{fiteq_w} 
\ee
For the real part of the fluidity, which was the quantity investigated in Ref. \onlinecite{bie17}, there is no contribution from the  plateau parameter. Referring to the simulation data in \fig{fig3}, one can easily imagine the plateau parameter $\beta$ to be non-universal, leading to a universal inherent MSD but a non-universal full MSD. Via \eq{fiteq_w} this would lead to a universal real part of the fluidity, but a non-universal imaginary part. 
This may explain why fluidity universality was not noted before: the more commonly studied frequency-dependent viscosity, $\eta(\omega)=1/F(\omega)$, is not universal.

\begin{figure}[htbp!]
	\includegraphics[width=8cm]{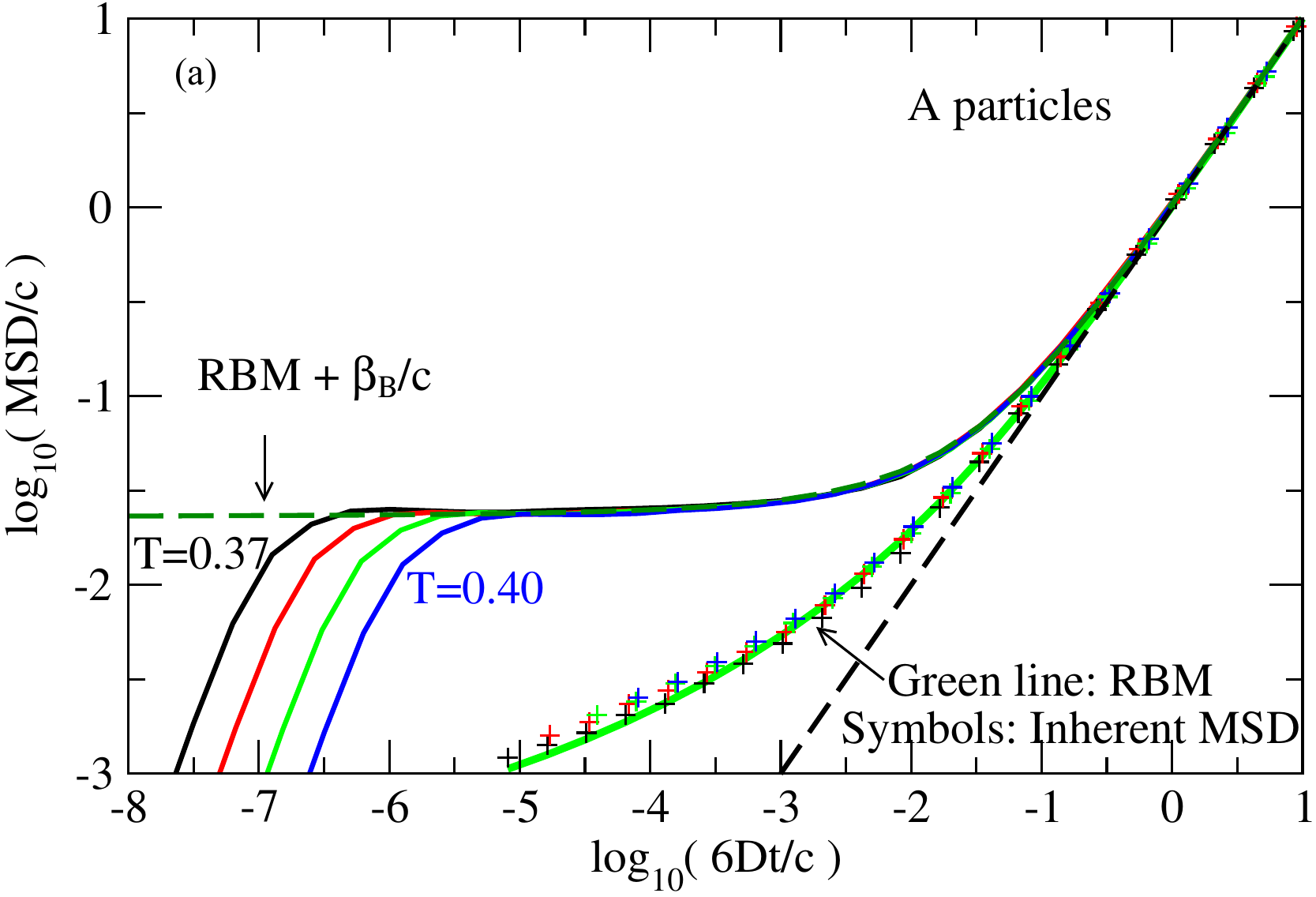}
	\includegraphics[width=8cm]{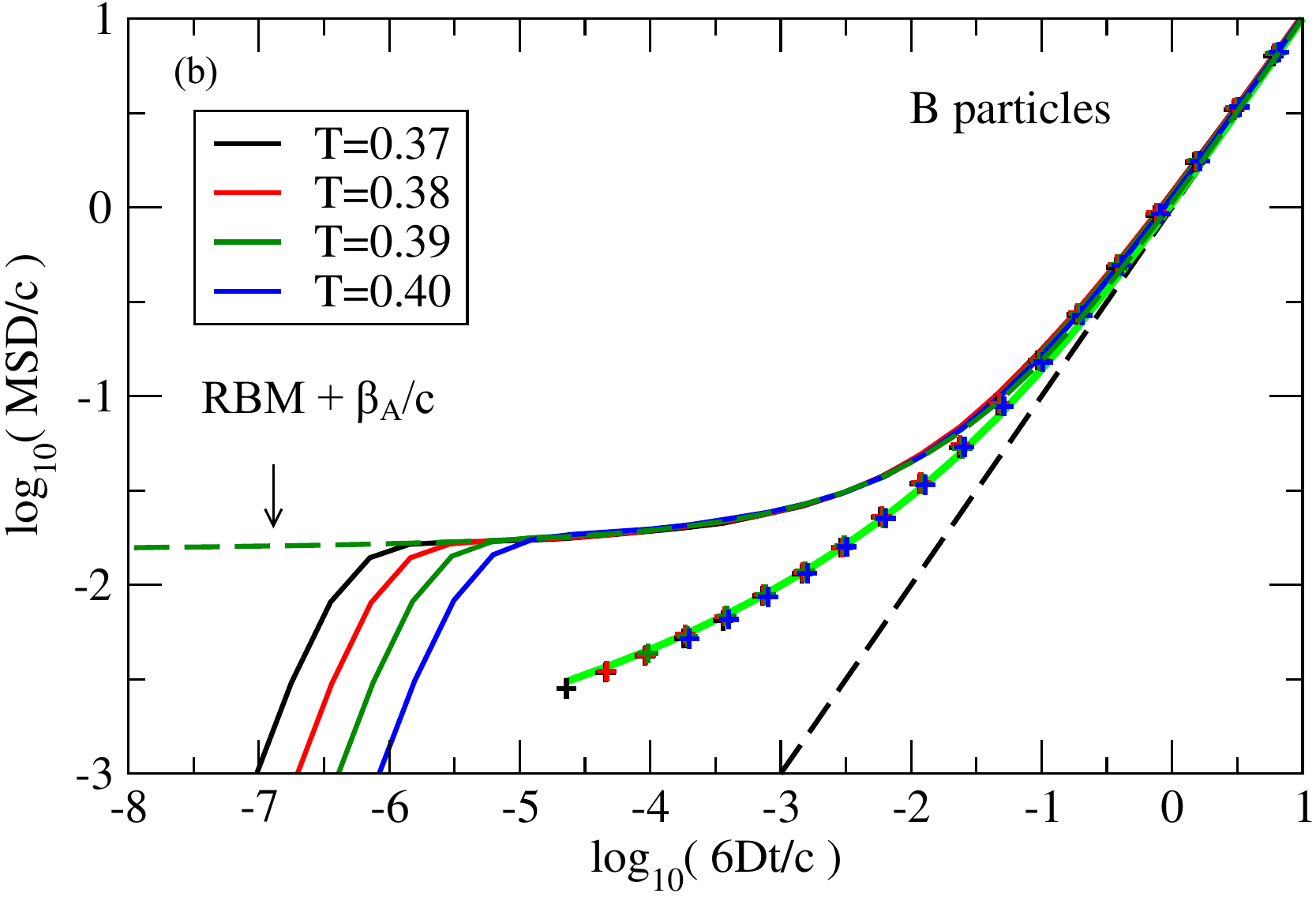}
	\caption{\label{fig4} Testing the RBM separately for each particle species with (a) showing the A particle MSD versus the RBM prediction and (b) showing the same for the B particles.}
\end{figure}  

Why do particles in a viscous liquid move like in a disordered solid? The liquid is disordered, of course, but the more or less random potential-energy landscape experienced by any given particle changes with time. This argument refers to three dimensions. Taking a more abstract approach, it has been argued that complexity may be replaced by randomness in the high-dimensional configuration space \cite{wol92}. This is similar to the philosophy of statistical mechanics, which ignores the extreme complexity of a many-body system and models it by a probability distribution. This way of thinking about the problem addresses also the challenge that a given particle does not experience a frozen landscape. \Fig{fig3} showed that the all-particle MSD is fitted well by the RBM prediction. \Fig{fig4}(a) shows results for the same analysis restricted to the A particles, while (b) shows it for the B particles. The fit to the RBM is not as good for the A particles as for all particles (\fig{fig3}), indicating that the dynamics of the A particles by themselves is not accurately modeled by the RBM. Interestingly, the B particles are fitted better by the RBM. The B particles are smaller than the A particles and they move faster, with a characteristic time for transition to diffusive dynamics that is about 1/10 of that of the A particles. This means that the B particles to a higher extent than the A particles move in a frozen landscape. Because the best fit to the RBM is found for the all-particle MSD, we speculate that the ``correct'' explanation of why the RBM works should refer to the high-dimensional configuration space, not to 3d space. 

Our findings focus on the shape of the MSD and do not have direct consequences for understanding the relaxation time's temperature dependence as modeled in, e.g., the Adam-Gibbs, random-first-order-transition, or shoving models \cite{dyr06}. The latter views the metastable equilibrium supercooled liquid as behaving more like ``a solid that flows'' than like an ordinary liquid \cite{dyr06}; the fact that particle motion in a highly viscous liquid in the present work has been shown to be much like particle motion in an disordered solid is in line with this view.

In summary, we have shown that the MSD of the KA model modified to avoid crystallization follows the zero-shape-parameter RBM prediction. This is consistent with the experimental findings of Ref. \onlinecite{bie17} that are also well described by the RBM. Our results leaves important questions for future works: Why do the mBLJ particles move as if they were hopping in a random solid with identical energy minima? How can one justify using a Stokes-Einstein argument for converting MSD to the frequency-dependent fluidity? Finally and perhaps most importantly: How general are our findings?

  \begin{acknowledgements}
  This work was supported by grants 00016515 and 00023189 from the VILLUM Foundation. 
\end {acknowledgements}

  \section{Appendix}

\begin{figure}
\begin{center} 
 \includegraphics[width=8cm]{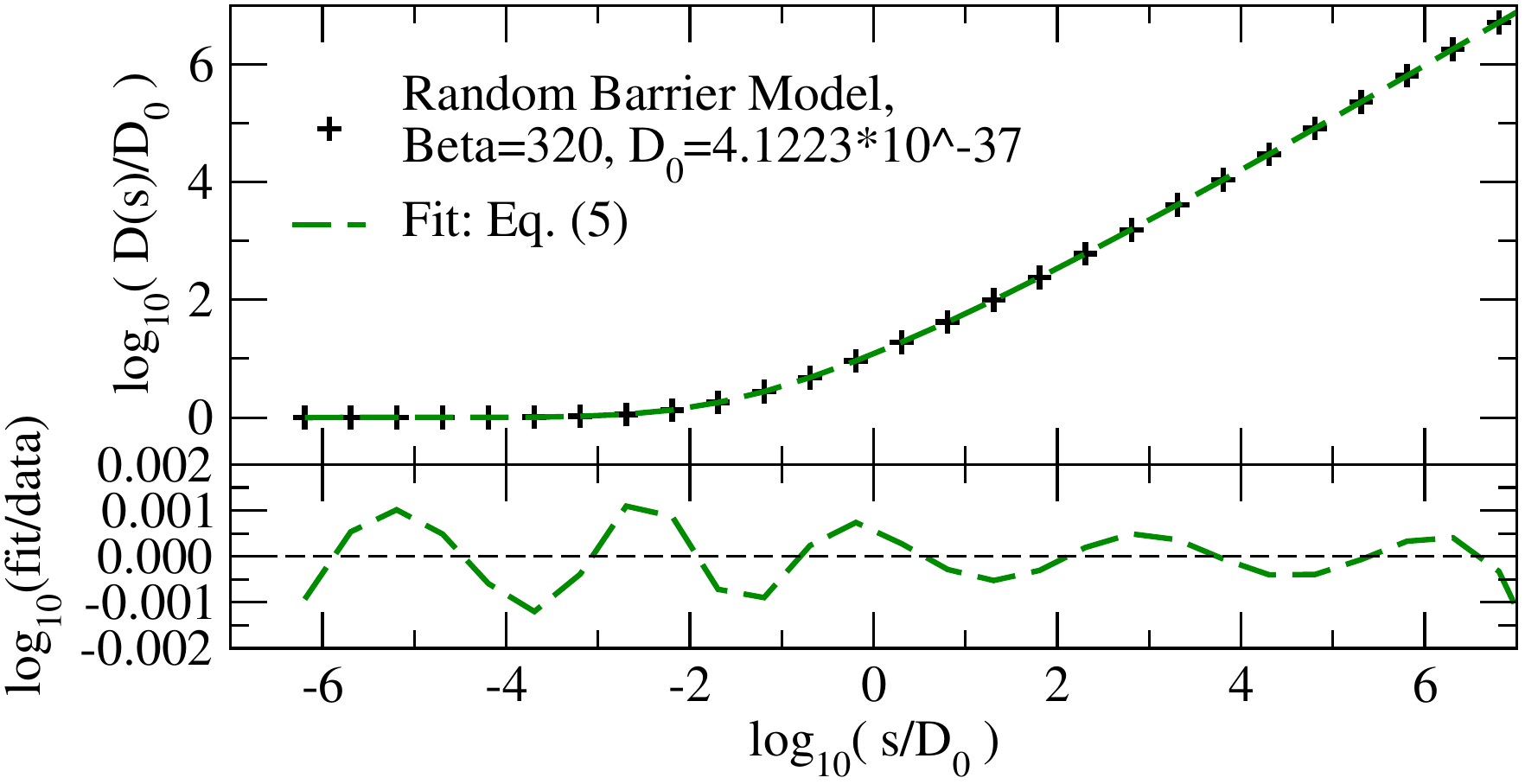}
\end{center}
\caption{Symbols: Numerical solution of the frequency dependent diffusion coefficent $D(s)$ for the Random Barrier Model at $\beta=320$, where $s$ is the real Laplace frequency, the Box distribution was used for the energy-barriers, and $\beta$ is the inverse temperature times the maximum barrier. Green dashed line are the fit according to Eq. (\ref{eq:FitRBMs}). The lower panel shows the deviation between fit and numerical data.
} \label{Sig320RBM}
\end{figure}

  The Random Barrier Model (RBM) was solved in the real-Laplace-frequency domain on a cubic lattice with a box distribution of energy barriers \cite{sch08}. The frequency-dependent diffusion coefficient $D(s)$ for $\beta =320$ is shown in \fig{Sig320RBM} as symbols \cite{sch08}, where $\beta$ is the maximum barrier over the temperature. At low temperatures the shape of $D(s)$ for the RBM becomes universal, i.e., independent of temperature and energy barrier distribution. In the frequency range shown, the  data in \fig{Sig320RBM} is a good representation of the universal RBM prediction.

To facilitate transformation from the real-Laplace-frequency domain to the time domain over the many decades involved, we fit to the following empirical fitting function in which $\tilde s \equiv s/D(s)$
 
\begin{equation}
  \frac{D(\tilde s)}{D_0} = 1 + \sum_{j=1}^{10} a_j\tilde s^{j/10} \label{eq:FitRBMs}
\end{equation}
The fitting was performed after taking the logarithm of both the numerical data and the fitting function, resulting in the following parameters $(a_1, ..., a_{10})$ =
(-1.1914,  11.2368, -34.2903, 26.6019, 47.0002, -96.2905, 77.4671, -27.0535, 7.72535, -0.178844). The resulting fit is shown as the green dashed line in \fig{Sig320RBM} and the corresponding error is shown in the lower panel.

Using $D(s)\equiv (s^2/6)\int_{0}^{\infty}\langle\Delta r^2(t)\rangle\exp(-st)dt$ (see the main text), \eq{eq:FitRBMs} implies

\begin{equation}
  \langle\Delta r^2(\tilde t)\rangle = 6D_0\tilde t \left(1 + \sum_{j=1}^{10} \frac{a_j {\tilde t}^{1-j/10}}{\Gamma(2-j/10)}\right) \label{eq:Fit}
\end{equation}
This is the equation used to represent the RBM in \fig{fig3} and \fig{fig4}, in which a further empirical rescaling of $\tilde t$ is introduced.

\end{document}